\documentstyle[12pt]{article}

\advance\voffset by -1.5cm
\advance\hoffset by -1.5cm
\title{\mbox{\large{Cosmological Consequences of
Superconducting String Networks}}} 
\author{
\mbox{\large{Konstantinos Dimopoulos}}
\thanks{e-mail: K.Dimopoulos@damtp.cam.ac.uk}
\vspace{0.2cm}\\
\mbox{\large{and}}
\vspace{0.2cm}\\
\mbox{\large{Anne--Christine Davis}}
\thanks{e-mail: A.C.Davis@damtp.cam.ac.uk}
\vspace{0.4cm}\\
\mbox{\normalsize{\em Department of Applied Mathematics and
Theoretical Physics,}}\\
\mbox{\normalsize{\em University of Cambridge, Silver Street,}}\\
\mbox{\normalsize{\em Cambridge, CB3 9EW, U.K.}}
}
\begin{document}
\begin{titlepage}
\maketitle
\begin{abstract}

We consider the cosmological consequences of a network of superconducting
cosmic strings. For strong enough current the period of friction domination
never ends. Instead a plasma scaling solution is reached. We demonstrate 
that this gives rise to a very different cosmology than the usual horizon
scaling solution. In particular the string network gives rise to a distinct
imprint on the microwave sky, giving non-Gaussian features on much smaller
angular scales. It also gives rise to a filament structure in string wakes.
Because of the presence of the string magnetocylinder, the string magnetic 
field cannot create a primordial magnetic field. Similarly, it evades
nucleosynthesis constraints. We also show that strings formed at the 
supersymmetry breaking scale can create the required baryon asymmetry 
of the universe.

\end{abstract}
\vspace{1cm}
\flushright{DAMTP-98-15}
\end{titlepage}

{\bf 1. Introduction}

\bigskip

The microphysics of cosmic strings has received considerable attention.
In particular, Witten \cite{witt} showed that cosmic strings become
superconducting as a result of boson condensates or fermion zero modes
in the string core. Such strings are capable of carrying a sizeable 
current, with the maximum current being about  \mbox{$10^{20}A$}
for a grand unified scale string. Inevitably, such currents have
cosmological and astrophysical \cite{astro} consequences. The
consequences for emission of synchrotron radiation  \cite{field} and
for high energy $\gamma$-rays \cite{grays}
have been explored. 
However, all these studies have assumed that the evolution 
of a network of superconducting strings is similar to that of ordinary
strings. 

Early studies using both analytic \cite{MB} and numerical
techniques \cite{num}
showed that the string evolution was indeed similar
to that of ordinary cosmic strings. However, these studies neglected 
the very early times when the string is interacting strongly with the
surrounding plasma. As shown recently in \cite{ours}, for strong enough 
currents, this friction dominated period may never 
end. In this case the network reaches the so-called plasma--scaling 
solution, where the density of strings may be much larger than that of the
usual horizon--scaling string networks. In this case the strings are 
highly tangled and move rather slowly. Also considering the 
attractive gravitational fields generated by superconducting strings 
\cite{mine}, it is evident that,
compared with ordinary cosmic strings, a network of superconducting 
strings could have very different cosmological implications on the 
matter and radiation content of the early universe.

In this letter we briefly explore some of the most important 
cosmological consequences of a plasma--scaling superconducting string 
network. Assuming a constant string current, we first discuss the 
characteristics of the plasma--scaling solution. We then describe the 
string spacetime and its implications on the surrounding particles and 
radiation. Afterwards, we explore the implications of the network 
regarding the Cosmic Microwave Background Radiation (CMBR) anisotropies 
and the formation of the Large Scale Structure (LSS) of the Universe. We 
also discuss any effects a plasma--scaling string network may have on Big 
Bang Nucleosynthesis (BBN) and whether it can generate a Primordial 
Magnetic Field (PMF) sufficient enough to account for the currently 
observed magnetic fields of the galaxies. 

Finally, we apply our results to a recently proposed baryogenesis mechanism
with superconducting cosmic strings \cite{BR}. This mechanism uses strings
formed at the supersymmetry breaking scale. With our plasma scaling solution
this mechanism produces a sufficiently strong baryon asymmetry to account
for nucleosynthesis. This improves the result of \cite{BR}, where
vorton domination was needed to obtain enough baryon asymmetry. In what 
follows, unless stated otherwise, we use natural units (\mbox{$\hbar=c=1$}).

\bigskip

{\bf 2. Friction and plasma--scaling}

\nopagebreak[4]

\bigskip

\nopagebreak[4]

After the formation of the string network,
curves and wiggles on the strings tend to untangle due to string
tension, which results in oscillations of the curved string segments
on scales smaller than the causal horizon and larger than their
curvature radii. Friction dissipates the energy of these oscillations
and leads to their gradual damping \cite{kibb}. Thus, the strings become 
smooth on larger and larger scales with their curvature radius $R$ 
growing accordingly. 

Friction on a cosmic string is caused by the interaction
of the string fields with the plasma particles. 
As shown in \cite{ours}, for string currents $J$ larger than a 
critical value $J_{c}$, the friction force is determined by the 
string magnetic field generated by the current.
This critical current is estimated as,

\begin{equation}
J_{c}\sim J_{max}\sqrt{LG\mu}
\label{Jc}
\end{equation}
where $\mu$ is the string mass per unit length,
\mbox{$G=m_{P}^{-2}$} is Newton's gravitational constant (with 
\mbox{$m_{P}=1.22\times 10^{19}GeV$} being the Planck mass) and
\mbox{$J_{max}\sim\sqrt{\mu/L}$} is the maximum acceptable string 
current, over which the string loses its superconducting properties 
\cite{witt}, with \mbox{$L\simeq\ln(\Lambda R)$} being 
the self-inductance of a string of radius $\Lambda^{-1}$ 
\cite{mine}.\footnote{In \cite{ours}, the factors of $L$ were omitted for 
simplicity. However, it is important to maintain them for quantitative 
calculations. Indeed, it can be easily shown that, for strings formed at 
the breaking of the Grand Unified Theory (GUT-strings) \mbox{$L\sim 
100$}.} 

For \mbox{$J\geq J_{c}$} friction prevents the network from reaching 
horizon--scaling \cite{ours}. Instead the strings are found to move with 
a more or less constant terminal velocity,

\begin{equation}
v\sim\sqrt{\frac{J_{c}}{J}}\sim
\left[\frac{G\mu}{\sqrt{GJ^{2}}}\right]^{1/2}\ll 1
\label{v}
\end{equation}

In this case the string network does satisfy a scaling solution 
(so-called plasma--scaling), which, however, may differ substantially 
from the usual horizon--scaling of ordinary strings. Indeed, a 
plasma--scaling network consists of slowly moving, highly tangled 
strings, with curvature radius and inter-string distance 
much smaller than the horizon, since \mbox{$R\sim vt\ll H^{-1}$}, where 
\mbox{$H\sim t^{-1}$} is the Hubble parameter.\footnote{Note that in the 
one-scale model the curvature radius and the inter-string distance of the 
string network are of similar magnitude.} Still, although denser, a 
plasma--scaling network is not in danger of dominating the overall energy 
density of the universe because, \mbox{$\rho_{s}/\rho\sim\sqrt{GJ^{2}}\ll 
1$}, where \mbox{$\rho_{s}\sim\mu/R^{2}$} is the energy density of the 
strings and \mbox{$\rho\sim 1/Gt^{2}$} is the energy density of the 
universe with $t$ being the cosmic time.

It should be noted here that $J$ is the value of the local, coherent
current on the string, which generates the string magnetocylinder and
determines the frictional cross-section between the string and the plasma.
$J$ should be thought as a free parameter, which, however, is expected to
assume large values \cite{witt} either
directly at current generation or through subsequent interactions with
weak primordial magnetic fields \cite{Miji}
On large scales the orientation of the string current is expected
to be the stochastic, or Kibble current \cite{edmarkneil}. The string current 
is dynamically conserved, which, in view of the growth in the curvature radius during string network evolution and 
causal interaction of different current patches on the string, results
in the local value $J$ remaining approximately constant \cite{ours}.
We note that superconducting strings in the friction dominated era have also
been considered in \cite{carlos}. However, the much smaller rms current was 
used rather than the local, coherent current. As a consequence the effect of 
the magnetocylinder, and thus the string interaction with the plasma,
was not included. This resulted in ref \cite{carlos} concluding that frictional
effects would be small in contrast to our results \cite{ours}.

\bigskip

{\bf 3. The string gravitational field}

\nopagebreak[4]

\bigskip

\nopagebreak[4]

The exact metric of the spacetime around a current carrying string was
first calculated by Moss and Poletti \cite{moss}. The implications of this
spacetime on test particles and light rays was also investigated 
\cite{demianski}, reaching similar conclusions as ref \cite{moss}. 
However, it was Linet \cite{linet} who first attempted to explore the 
gravitational properties of a superconducting string system in a more 
realistic way, by considering only first order terms in $G$. 
Linet demonstrated that this was fully consistent with the 
original, exact solution. A more thorough study of the linearised spacetime 
of a superconducting string \cite{HK} arrived at similar conclusions.
Finally, these results were extended \cite{peter} by also considering higher 
order terms in the gauge coupling. 
However, in all the above work the importance
of self-inductance effects on the string spacetime has not been fully
appreciated. Taking these into account \cite{mine} has shown that the 
perpendicular to the string geometry is described by,

\begin{equation}
ds^{2}_{\perp}=(1+2\Phi)[-dt^{2}+dr^{2}+(1-\delta/\pi)r^{2}d\theta^{2}]
\label{ds}
\end{equation}
where \mbox{$\Phi\sim L(GJ^{2})\ln(\Lambda r)$} is the attractive 
gravitational potential and $\delta$ is the deficit angle estimated as 
\cite{mine},

\begin{equation}
\delta\simeq 8\pi G(\mu+LJ^{2})
\label{delta}
\end{equation}

Therefore, the existence of a current on the string generates an attractive 
gravitational field. This field along with the conical form of the 
spacetime affects the surrounding particles while the string moves in 
the plasma. In \cite{mine} it is shown that the velocity boost felt by 
the particles towards the perpendicular direction to the string motion is,

\begin{equation}
u=8\pi G\mu v\gamma+8\pi GJ^{2}L\left(v\gamma+\frac{1}{v\gamma}\right)
\label{u}
\end{equation}
where \mbox{$\gamma^{-1}=\sqrt{1-v^{2}}$} is the Lorentz factor. In the 
above the first two terms are due to the deficit angle whereas the last 
term is due to the attractive gravitational field. The gravitational 
field dominates for \mbox{$J\geq J_{G}$} where,\footnote{In can be easily 
verified that, for realistic values of the parameters, 
\mbox{$J_{G}>J_{c}$}.} 

\begin{equation}
J_{G}\sim J_{max}(LG\mu)^{1/6}
\label{JG}
\end{equation}

The effect of the existence of strong string currents on the string 
spacetime morphology and on the characteristics of the string network 
scaling solution is expected to reflect itself on the numerous
cosmological implications of strings, in particular the CMBR anisotropies 
and the formation of the LSS of the universe. 

\bigskip

{\bf 4. Anisotropies on the microwave sky}

\nopagebreak[4]

\bigskip

\nopagebreak[4]

The root mean square (rms) CMBR temperature anisotropies generated by 
cosmic strings may be estimated as \cite{leandros},

\begin{equation}
\left(\frac{\Delta T}{T}\right)_{rms}
\simeq\sqrt{N}
\left(\frac{\Delta T}{T}\right)_{S}
\label{DTrms}
\end{equation}
where \mbox{$N\sim (HR)^{-2}$} is the number of strings inside a 
horizon volume (see \cite{ours}) and

\begin{equation}
\left(\frac{\Delta T}{T}\right)_{S}\simeq\delta v\gamma
\label{DTs}
\end{equation}
is the anisotropy generated by a single string \cite{KS}. 
Thus, from (\ref{delta}) and the above, the rms anisotropy 
generated by a network of superconducting strings is \cite{ours},

\begin{equation}
\left(\frac{\Delta T}{T}\right)_{rms}\simeq
\delta\simeq 8\pi G(\mu+LJ^{2})
\label{DT}
\end{equation}

The above suggests that the rms effect of the string spacetime on 
radiation does not depend on the gravitational field of the strings. 
This is to be expected since, for radiation, \mbox{$ds_{\perp}=0$} 
and (\ref{ds}) suggests that the prefactor \mbox{$(1+2\Phi)$}  
cannot influence the shape of the null geodesics. Moreover, because 
\mbox{$LJ^{2}\leq\mu$} the magnitude of the rms temperature 
anisotropies is little affected by the string current. However, 
in terms of the stochastic nature of the anisotropy distribution,
a plasma--scaling string network may produce a distinct imprint on
the microwave sky, due to the larger number of strings per horizon. 
Since the string network is denser one possible effect is to shift 
the position of the Doppler peak to smaller values of $l$.

Indeed, the distribution of CMBR temperature anisotropies generated 
by a horizon--scaling network of ordinary strings is expected to be
non-Gaussian over angular scales smaller than 
\mbox{$(\Delta\vartheta)_{0}\sim 1^{\circ}$}, which 
corresponds to the angular scale of the horizon at the time of last
scattering \cite{eust}. However, as the inter-string distance is much
smaller in a plasma-scaling string network, one would expect to discover
non-Gaussian signatures only on angular scales smaller than, 

\begin{equation}
\Delta\vartheta\sim\frac{R}{H^{-1}}(\Delta\vartheta)_{0}\sim 
v(^{\circ})\ll 1^{\circ}
\end{equation}

For GUT-strings the rms anisotropy 
is \mbox{$(\frac{\Delta T}{T})_{rms}\simeq 8\pi G\mu\sim 10^{-5}$}, 
in good agreement with the observations. In this case, 
it is easy to see that,
for maximum string current, the Gaussianity of the distribution 
appears over angular scales less than 0.1$^\circ$ as,

\begin{equation}
v(J_{max})\sim (LG\mu)^{1/4}
\end{equation}

However, in order to ascertain
the full string predictions for the CMBR anisotropies large computer 
simulations are necessary. Whilst early simulations produced disappointing
results suggesting the lack of a Doppler peak \cite{neil} the most recent
simulation suggests that local cosmic strings can account for the 
observed CMBR \cite{joao}. It is likely that our denser network will 
have more power on small scales since its scaling distance is smaller. 
A full scale numerical simulation is required to compute the exact scale
of the peak in the power spectrum. This is the subject of a future 
investigation.

\bigskip

{\bf 5. Large scale structure overdensities}

\nopagebreak[4]

\bigskip

\nopagebreak[4]

The angular deficit of the string spacetime and the attractive 
gravitational field generate two overlapping streams of matter 
behind a moving string. This is because of the relative boost, $u$,
felt by the plasma particles towards the string trail. Thus, 
the matter overdensity generated by a moving string may be 
estimated as, \mbox{$\delta\rho=\beta\rho$}, where 
\mbox{$0\!<\!\beta\!\leq\!1$} is determined by the 
fraction of the matter streams that remain inside the string 
wake, rather than dissipating into the inter-string space. 
The $\beta$ factor is strongly related to the nature of the dark
matter of the universe. For baryonic or Cold Dark Matter (CDM)
\mbox{$\beta\simeq 1$} as almost all the overdensity is contained 
inside the string wake. For Hot Dark Matter (HDM) though, $\beta$ 
can be substantially smaller as an important fraction of the 
overdensity diffuses away due to free streaming effects 
\cite{book}.

The length of a string wake is \mbox{$l(t)\sim vt$} and its 
thickness is \mbox{$d(t)\sim ut$}, where $u$ is given by 
(\ref{u}). Thus, the linear mass overdensity of the wake is
\mbox{$\delta\mu=(\delta\rho)dl\simeq\beta\rho uvt^{2}$}. 
Therefore, the total overdensity of a string wake is,

\begin{equation}
\left(\frac{\delta\rho}{\rho}\right)\simeq
\frac{1}{\rho}\frac{\delta\mu}{R^{2}}\sim\beta\frac{u}{v}
\label{drho}
\end{equation}

For currents smaller than $J_{G}$ the boost $u$ is determined
by the deficit angle terms in (\ref{u}). In this case it easy 
to see that,

\begin{equation}
\left(\frac{\delta\rho}{\rho}\right)_{J<J_{G}}\simeq\beta\delta
\label{drho1}
\end{equation}
where we have taken \mbox{$\gamma\simeq 1$} since the 
coherent motion of the strings is never expected to be 
ultrarelativistic \cite{ours}. 

The above estimate is not very different from the case of ordinary 
strings, which again is due to the fact that the deficit angle is 
largely insensitive to the string current. However, when the 
gravitational field becomes important, the situation is drastically 
changed.

For very strong currents the gravitational attraction term
dominates in (\ref{u}). In this case, using also (\ref{v}) one finds,

\begin{equation}
\left(\frac{\delta\rho}{\rho}\right)_{J\geq J_{G}}\simeq 
\beta(8\pi L)\frac{(GJ^{2})^{3/2}}{G\mu}
\label{drho2}
\end{equation}

For maximum current the above becomes,

\begin{equation}
\left(\frac{\delta\rho}{\rho}\right)_{J=J_{max}}\simeq 
\beta(\frac{8\pi}{\sqrt{L}})\sqrt{G\mu}
\label{drhomax}
\end{equation}

Observations of the galaxy correlation function of the LSS
suggest that \mbox{$(\frac{\delta\rho}{\rho})_{obs}\sim 10^{-5}$} 
\cite{book}. Therefore, for GUT-strings with weak currents
there is reasonable agreement for CDM models with 
\mbox{$\beta\simeq 1$}. However, for strong currents HDM or MDM 
(Mixed Dark Matter) models are preferable. Indeed, for maximum 
current (\ref{drhomax}) suggests that \mbox{$\beta\leq 0.1$}. 
In general, from the comparison with observation it can be shown 
that, for GUT-strings with \mbox{$J\geq J_{G}$} one requires,

\begin{equation}
J\leq \beta^{-1/3}J_{G}
\end{equation}

The above constraints may be somewhat strengthened if the 
distribution of dark matter is smoother than the distribution of 
the galaxies. Indeed, it is believed that the observed galactic 
distribution, which is used in order to estimate the density 
perturbations in the universe today, represents only the peaks in 
the actual density distribution of the dark matter. It is, thus, 
believed that the overall density perturbation of the universe 
relates to that observed as, \mbox{$(\frac{\delta\rho}{\rho})_{obs}=
b(\frac{\delta\rho}{\rho})$}, where \mbox{$b\geq 1$} is the
so-called bias factor \cite{book}. From the above it is evident 
that this factor may be included in $\beta$ and so the form of our 
results remains unaffected. Also, since $b$ is expected to be of order
unity, the quantitative estimates remain reliable.

Apart from the magnitude of the overdensities a plasma--scaling
string network may generate LSS morphologically different from
the one due to ordinary strings. Indeed, the slow moving strings of
a friction dominated network would produce filaments, rather than thin
wakes. It is possible that these filaments are thickened by gravitational
effects. The distribution of these filaments would be denser due to
the smaller inter-string distance of the network. This is rather 
unfortunate as the spectrum of density distributions would lose power
on large scales, a problem already present for ordinary strings. 
Thus,
one could argue that plasma--scaling superconducting string networks
are not sufficient to explain the overall LSS, and some other 
density perturbation mechanism is required to seed the structure on 
very large scales. If such a mechanism exists then the smaller 
filamentary structure generated by the strings could be swept inside 
the `pancakes' of the larger, horizon--sized density perturbations. 
Such structures, i.e. embedded filaments on large walls are indeed 
observed \cite{KT}. However, a full numerical simulation is required
to investigate this and is the subject of future investigation.
If the numerical simulations confirm these tentative conclusions then
possible mechanism to generate both cosmic strings 
and large scale density perturbations could be Hybrid Inflation 
\cite{hybrid}.

\bigskip

{\bf 6. Nucleosynthesis and galactic magnetic fields}

\nopagebreak[4]

\bigskip

\nopagebreak[4]

In an early work of Butler and Malaney \cite{MB} it was suggested 
that the existence of a network of electrically charged current carrying 
strings may seriously disturb Big Bang Nucleosynthesis (BBN). Their
argument was based on the fact that a current carrying string generates
a Biot-Savart magnetic field which results in the creation of a 
magnetocylinder around the string core \cite{field}\cite{book}\cite{ours}.
This magnetocylinder is impenetratable to charged plasma particles but not 
to single neutrons. Thus, in \cite{MB} it was argued that inside the trail 
of the moving strings an overdensity of neutrons would be generated that 
may affect the rate of BBN's reactions and the abundance of the resulting 
elements. 

However, it can be easily shown that this is not actually the case. 
Firstly, the charged plasma particles that are pushed away on the border
of the magnetocylinder follow the magnetic field lines in
a similar way that the solar wind is directed towards the 
Earth's magnetic poles by the Earth's Magnetosphere. Thus, the orbits of 
the charged plasma particles trace the surface of the magnetocylinder and, 
therefore, are expected to be sucked back into the trail of the string 
after the string has passed. Moreover, not only does the charged plasma
close behind the string magnetocylinder but some of it may even penetrate 
it from the back as discussed in \cite{book}. 

Another argument against the disastrous implications of \cite{MB} is 
due to purely geometrical facts. The dimensions of the string 
magnetocylinder are determined by the pressure balance between the plasma
and the string magnetic field as \cite{field}\cite{ours},

\begin{equation}
r_{s}\sim \frac{J}{\sqrt{\rho}\,v}
\end{equation}
where $v$ is the string velocity. The plasma--scaling solution suggests
that, inside a volume $\sim R^{3}$ one would expect only about one string 
segment of length $\sim R$.  This segment is expected to sweep, while 
moving, a volume \mbox{$\Delta V\sim r_{s}\times R\times v\Delta t$}. 
Thus, using \mbox{$R\sim vt$}, the fraction of volume
traced by string magnetocylinders per Hubble time $t$ is,

\begin{equation}
\frac{\Delta V}{R^{3}}\sim\frac{r_{s}}{R}\sim\frac{J^{2}}{\mu}\leq 
L^{-1}\sim 10^{-2}
\label{frac}
\end{equation}
Therefore, since the duration $\Delta t_{BBN}$
of BBN  is no more 
than about a hundred Hubble times \cite{KT}, an arbitrary point in space 
may be swept by a string magnetocylinder at most once or twice. Such an 
encounter would last about 
\mbox{$\Delta t\sim r_{s}/v\sim L^{-1}t\ll\Delta t_{BBN}$} Thus, one 
would expect that any effect that such an event may have on BBN's 
processes would be insignificant. 

It would be misleading to believe that BBN's processes could be disturbed
by the long--range Biot-Savart string magnetic field. Indeed, the field is,
in fact, not expected to extend beyond the border of the string 
magnetocylinder because the charged plasma particles, while travelling on
the magnetocylinder's border, generate a surface current of 
opposite orientation to the string current \cite{ours}. As a result, {\em 
the string magnetic field is cancelled outside the magnetocylinder}.

For the same reason the string magnetic field cannot be involved in any
astrophysical processes, since, being contained inside the 
magnetocylinders of the strings, it never really comes into contact with
the cosmic plasma. Thus, such a field cannot freeze into the plasma and 
be in any way directly responsible for seeding the galactic magnetic fields.
However, as shown in \cite{mine}, superconducting cosmic strings may 
efficiently generate a primordial magnetic field indirectly, through 
dynamical friction. Such a field may be strong and coherent enough to 
easily trigger the galactic dynamo and generate the observed galactic 
magnetic fields. Also, it can be shown that the the plasma vorticity 
generated by the string motion and gravitational pull may be contribute to 
the fragmentation process of galaxy formation as the scale of the 
spinning plasma volumes compares to the protogalactic scale  
before gravitational collapse \cite{mine}. 

\bigskip

{\bf 7. Baryogenesis}

\nopagebreak[4]

\bigskip

\nopagebreak[4]

In a recent work Brandenberger and Riotto \cite{BR} have suggested a new
mechanism for explaining the baryon asymmetry in the Universe, involving
superconducting cosmic strings. Their model considers a cosmic string network 
formed at the breaking of supersymmetry at temperatures of order  
\mbox{$10^{2}TeV$}. Charged sleptons and squarks condense in the string core,
resulting in bosonic superconductivity. CP-violating interactions
during the string network formation period may result in the confinement 
of a non-zero net baryon number inside the string core, which would be 
preserved due to dynamical and topological current conservation until after
the electroweak phase transition, when it could be released through string 
loop decay without being erased by sphaleron processes. In their 
treatment Brandenberger and Riotto have shown that the confined baryon 
number is released during the friction domination period of the string
network. However, they did not take into account 
the effects of excessive friction on the string evolution due to 
the existence of a magnetocylinder around the string core. 

As discussed in previous sections, for large currents 
a network of superconducting strings remains always friction dominated. 
Thus, such a friction--scaling string network would be more tangled and 
denser, which would imply that the captured baryon number density may be
larger than the original considerations of Brandenberger and Riotto. 
Following the reasoning of \cite{BR} we calculate the baryon
number density generated by a network of superconducting strings carrying
maximum current \mbox{$J\sim 10^{2}TeV$}.

Using the one--scale model, the number density of loops created per 
unit time is given by 
\cite{book},

\begin{equation}
\frac{dn}{dt}=\nu R^{-4}\frac{dR}{dt}\sim\frac{v}{R^{4}}
\label{dn}
\end{equation}
where $\nu$ is a numerical factor of order unity. Each of these loops contains
a net baryon number trapped inside the strings 
at the time of their formation $t_{0}$. If $Q$ is the baryon charge per 
unit length, and  \mbox{$Q\sim\sqrt{\mu}$} \cite{BR} then the charge per 
correlation length at the network formation 
is, \mbox{$Q_{0}=QR_{0}$}, where \mbox{$R_{0}\sim (\lambda\sqrt{\mu})^{-1}$} 
is the initial correlation length \cite{berger} with $\lambda$ being the
self--coupling of the string vortex field. The 
baryon charge on larger string segments can be estimated using a random walk.
Thus, when a loop of radius $R(t)$ is formed one would expect it to contain 
baryon charge of order

\begin{equation}
Q_{R}(t)\sim
\left[\frac{R(t)}{\left(\frac{a(t)}{a(t\!_{0}\!)}R_{0}\right)}\right]^{1/2}
Q_{0}
\label{QR}
\end{equation}
where \mbox{$a(t_{0})\propto t^{1/2}$} is the scale factor of the Universe 
and we have included the conformal stretching of the strings. When 
the loop decays a fraction \mbox{$\epsilon\leq 1$} of the captured baryon 
charge is released as a net baryon number, 
\mbox{$\Delta n_{B}=\epsilon Q_{R}$}, where $\epsilon$ is determined by 
the rates of CP-violating processes \cite{BR}.

The total baryon number density generated by loop decay is easily estimated
as \cite{BR},

\begin{equation}
n_{B}(t)=\int_{t_{i}}^{t}dt'\epsilon Q_{R}(t')\frac{dn}{dt}(t')
\left(\frac{t'}{t}\right)^{3/2}
\label{nBint}
\end{equation}
where the final factor is due to cosmological redshift and
$t_{i}$ is the earliest time, when loops that contribute to the baryon 
number density are formed. Using (\ref{dn}) and (\ref{QR}) equation 
(\ref{nBint}) gives,

\begin{equation}
n_{B}\simeq\frac{5}{4}\epsilon\nu Q_{0}R_{0}^{-3}
\left(\frac{t_{0}}{t}\right)^{3/2}
\left(\frac{t_{0}}{t_{i}}\right)^{5/4}
\label{nB}
\end{equation}
This gives,

\begin{equation}
\frac{n_{B}}{s}\sim\epsilon\nu Q_{0}\lambda^{3}g_{*}^{-1}
\left(\frac{T_{i}}{T_{0}}\right)^{5/2}
\label{nB/s}
\end{equation}
where $s$ is the entropy density of the Universe,
\mbox{$g_{*}\sim 10^{2}$} is the number of degrees 
of freedom and \mbox{$T_{i}=T(t_{i})$}.

In order to evaluate the above one needs to decide on the choice of
$T_{i}$. If the loops decay promptly they do so in less than a Hubble time
due to the efficient radiation emission \cite{BR}. In this case, at
earlier times than the electroweak phase transition, any
baryon number released is expected to be `washed-out' by sphaleron 
processes. Thus, only loops formed later than the time 
$t_{ew}$ of the transition may contribute to the net baryon number
density. Therefore, 
\mbox{$T_{i}\simeq T_{ew}\equiv T(t_{ew})\sim 10^{2}GeV$} and 
equation (\ref{nB/s}) becomes,

\begin{equation}
\frac{n_{B}}{s}\sim\epsilon\nu Q_{0}\lambda^{3}g_{*}^{-1}
\left(\frac{T_{ew}}{T_{0}}\right)^{5/2}
\label{n1}
\end{equation}
The above result differs substantially from the findings of Brandenberger 
and Riotto (equation (30) in \cite{BR}), by a factor 
\mbox{$(T_{ew}/T_{0})^{-2}\sim 10^{6}$}. Thus, the result of \cite{BR} 
underestimates the generated baryon asymmetry by a million times!
Taking \mbox{$Q\sim\sqrt{\mu}$} (i.e. \mbox{$Q_{0}\sim 1$}) it can be 
easily seen that the desired asymmetry \mbox{$n_{B}/s\sim 10^{-10}$} 
can be achieved with rather natural values of the parameters:
\mbox{$\epsilon\sim 10^{-1}$} and \mbox{$\lambda,\nu\sim 1$}.

However, if instead of collapsing the string loops form stable vortons, 
i.e. string rings stabilised by the angular momentum of the current carriers 
\cite{vortons}, the above situation is modified. Vortons may release their
baryon number if they ever become unstable and decay \cite{an}. In this case, 
vortons manage to preserve their baryon number throughout the period prior to 
the electroweak transition.
Thus, the resulting net baryon number density may receive 
contributions even from the time of network formation, i.e. 
\mbox{$T_{i}\simeq T_{0}$}. Consequently, (\ref{nB/s}) would give,

\begin{equation}
\frac{n_{B}}{s}\sim\epsilon\nu Q_{0}\lambda^{3}g_{*}^{-1}
\label{n2}
\end{equation}
which is identical with equation (31) of \cite{BR}. This is not 
surprising since, in this case, the integral of (\ref{nBint}) is 
dominated by the initial contribution at the time $t_{0}$ of formation of the 
string network, so that the subsequent frictional evolution of the strings
does not affect the results.

\bigskip

{\bf 8. Conclusions}

\nopagebreak[4]

\bigskip

\nopagebreak[4]

In conclusion, we have investigated the cosmological and astrophysical 
consequences of a plasma--scaling, charged-current carrying, open string 
network. 

We have shown that such a network 
would generate large scale structure with very different features
than the one produced by a horizon scaling network. Indeed, the slow
moving strings would create filaments instead of thin wakes, whose
separation distances would be much smaller than the horizon. 
This compounds the existing problem of structure formation with 
ordinary strings, due to the lack of power on large scales.
One way to overcome this is by considering hybrid models, which incorporate
inflation with cosmic strings (see for example \cite{hybrid}). In such 
models the large scale fluctuations could be generated by 
inflation and the string--produced filaments swept into the
horizon-sized `pancake' structures. As we have shown, the magnitude of 
such filamentary overdensities depends on the type of dark matter 
assumed, which gives upper bounds on the parameters, and
may provide a link between the bias factor and the string 
current. 

We have also found that
the imprint of the strings on the microwave sky would be Gaussian on
smaller angular scales than the horizon scale at decoupling. The scale 
of the non-Gaussian features depends on the string current and is related 
to the terminal velocity of the friction dominated strings.

Furthermore, 
we discussed possible effects of a plasma--scaling network on 
nucleosynthesis and showed that the latter is not seriously disturbed 
even for maximum string currents. We briefly considered the possibility of 
direct generation
of primordial magnetic fields by the string magnetic fields. Since such 
fields are shielded by the string magnetocylinder they are unable to 
freeze into the cosmic plasma and have any astrophysical effect.

We have shown that, regardless of the existence of stable vortons, 
superconducting cosmic strings that are formed at the breaking of 
supersymmetry are able to generate the observed baryon
asymmetry in the Universe. We showed that, in contrast to the claim 
in \cite{BR}, 
a friction--scaling string network can create the required baryon
number density even without the production of stable vortons, for 
rather natural values of the model parameters. This is due to the fact
that a friction--scaling network is much denser that a horizon--scaling
one, producing substantially more string loops, whose decay
eject sufficient baryon charge when decaying to account for the 
observed anisotropy. 

In overall, the plasma--scaling solution of electrically charged current 
carrying superconducting strings may result in a modified cosmic string 
cosmology.  Comparing this scenario with observations could provide insight 
into the microphysics of strings and the effect of cosmic string 
superconductivity.

\bigskip

{\bf Acknowledgements}

\bigskip

We would to thank Nathalie Deruelle for discussions. This work was 
supported in part by PPARC.

\end{document}